\documentstyle[epsfig,aps,pre,floats,tighten]{revtex}
\begin{document}
\draft
\title{II. Territory covered by $N$ random walkers on stochastic fractals. \\ The percolation aggregate}
\author{L. Acedo and S. B. Yuste}
\address{Departamento de F\'{\i}sica, Universidad  de  Extremadura,
E-06071 Badajoz, Spain}
\date{\today}
\maketitle
\begin{abstract}
The average number $S_N(t)$ of
distinct sites visited up to time $t$ by $N$ noninteracting random
walkers all starting from the same origin in a disordered  fractal is considered.  
This quantity $S_N(t)$ is the result of a double average: an average over random walks on a given lattice followed by an average over different realizations of the lattice. 
We show for two-dimensional percolation clusters at criticality (and conjecture for other stochastic fractals) 
that the distribution of the survival probability over these realizations is very broad in Euclidean space but very narrow in the chemical or topological space. 
This allows us to adapt the formalism developed for  Euclidean and deterministic fractal lattices to the chemical language, and an asymptotic series for $S_N(t)$ analogous to that found for the non-disordered media is proposed here. 
The main term is equal to the number of sites (volume) inside a
``hypersphere'' in the chemical space of radius $L [\ln (N)/c]^{1/v}$ where $L$ is the root-mean-square chemical displacement of a single random walker, and $ v $ and $c$ determine how fast $1-\Gamma_t(\ell)$ (the probability that a given site at chemical distance $\ell$ from the origin  is visited by a single random walker by time $t$) decays for large values of $\ell/L$: $1-\Gamma_t(\ell)\sim \exp[-c(\ell/L)^v]$.
The parameters appearing in the first two asymptotic terms of $S_N(t)$ are estimated by numerical simulation for the two-dimensional percolation cluster at criticality. The corresponding theoretical predictions 
are compared with simulation data, and the agreement is found to be very good.

\end{abstract}
\pacs{PACS numbers: 05.40.Fb,05.60.Cd,66.30.Dn}
\section{INTRODUCTION}
\label{sect_1}
In the preceding paper \cite{fractal1} we studied the problem of evaluating the average number $S_N(t)$ of distinct sites (or territory covered) by $N$ independent random walkers that,  all starting from the same site, diffuse on a deterministic fractal during time $t$. 
An answer to this problem  was given in Ref.\ \cite{fractal1} in terms of an asymptotic series completely analogous to that given in Refs.\ \cite{RC,PREeucl} for Euclidean substrates. 
In this present paper we turn our attention to the case of stochastic (or disordered)  fractal media and again propose the same kind of asymptotic solution, although, in contrast with all the previous cases, we now have to translate the procedure to the topological or chemical language.

Stochastic media are not constructed by the
iteration of an invariable rule, such as that used in deterministic
fractals, but are rather the result of a random process. Consequently, the concept of generator and rigorous self-similarity is absent and their fractal nature is recognized by the scaling of statistical quantities. 
Many natural objects share this statistical-fractal structure \cite{BundeHavlin,SHDBA} so that stochastic models seem to be more suitable to represent diffusion in real media. 
Of particular interest is the percolation model that has been used to characterize many disordered systems \cite{SHDBA,Bunde,Stauffer}. This model is constructed by filling a regular lattice with ``occupied'' sites with a certain probability. Nearest neighbor occupied sites
are supposed to be connected and form a series of clusters. At a certain critical concentration $p_c$ an infinite cluster appears, which is called the incipient percolation aggregate or percolation cluster at criticality. 

The quantity we are interested in, $S_N(t)$, is, for disordered media, the result of a double average: an average over the walks that the $N$ random walkers can perform over a given lattice, followed by an average over many (ideally, all) realizations of the random lattice. 
This situation leads to certain subtleties, absent in deterministic fractals, that require special treatment. 
In particular, $S_N(t)$ can be expressed  by
\begin{equation}
\label{SGSC}
S_N(t)=\left\langle\sum \, \left\{ 1 - \left[\Gamma_t({\bf r})\right]^N \right\}\right\rangle\; ,
\end{equation}
where the sum is over all the sites of each fractal lattice, 
$\Gamma_t({\bf r})$ is the survival probability, i.e., the probability that site ${\bf r}$ has not been visited by time $t$ by a single random walker starting from the origin, 
$\sum  \left\{ 1 - \left[\Gamma_t({\bf r})\right]^N \right\}$ represents the mean territory explored by the $N$ random walkers on a given lattice (the first average) \cite{Havlin}, and $\langle [\cdots]\rangle$ indicates that the average (the second average) of $[\cdots]$  has to be performed over all possible stochastic lattices compatible with the random generation rules. 
Equation (\ref{SGSC}) can be rewritten as
\begin{equation}
\label{SGSC1}
S_N(t)=\sum_{m=0}^{\infty} \left\langle  \sum_{i=1}^{n(m)}\, 
\left\{ 1 - \left[\Gamma_t({\bf r}_{m,i})\right]^N \right\} \right\rangle\; ,
\end{equation}
where ${\bf r}_{m,i}$ stands for the $i$-th site out of $n(m)$ that are separated from the origin by a Euclidean distance between $m \Delta r \equiv r_m $ and $(m+1) \Delta r$ with $\Delta r$ small (say, of the order of the lattice spacing). 
If $\Gamma_t({\bf r}_{m,i})$ is almost independent of $i$ and the lattice realization, i.e., if the fluctuations in the probability density $\Gamma_t({\bf r}_{m,i})$ follow a narrow distribution, then  one could approximate 
$\Gamma_t({\bf r}_{m,i}) \simeq \langle \Gamma_t({\bf r}_{m,i}) \rangle \equiv \Gamma_t(r_m)$,  and therefore estimate $S_N(t)$ by
\begin{equation}
\label{SGSC2}
S_N(t)=\sum_{m=0}^{\infty} \left\{1-\left[\Gamma_t(r_m)\right]^N \right\}\left\langle n(m) \right\rangle\; ,
\end{equation}
where $\left\langle n(m) \right\rangle$ is the average number of fractal sites separated from the origin by a distance bracketed by $r_m$ and $r_m+\Delta r$. 
This is essentially the starting relationship used (implicitly) by Havlin et al. \cite{Havlin}  to find that, for large $N$,
\begin{equation}
\label{SNt1}
S_N(t)\sim t^{d_s/2} (\ln N)^{d_f/u}
\end{equation}
in the non-trivial time regime (or regime II) \cite{fractal1}. 
Here $u=d_w/(d_w-1)$, $d_w$ is the anomalous diffusion exponent, $d_f$ is the fractal dimension of the substrate, and $d_s=2d_f/d_w$ is the spectral dimension.
However, the hypothesis leading to Eq.\ (\ref{SGSC2}) is in general false, as we will explicitly show in Sec.\ \ref{sect_2} by means of numerical simulations  of $\Gamma_t({\bf r}_{m,i})$ for the two-dimensional  percolation cluster at criticality.
Indeed, it is known \cite{RCBunde} that the fluctuations of the probability density $P(r,t)$ of random walks (also called the propagator or Green's  function), which is a statistical quantity closely related to the survival probability, exhibits a broad logarithmic distribution for some random fractals such as percolation clusters and self-avoiding walks.
Bunde {\em et al.} \cite{RCBunde} have found that the quantity
 $\langle P(r,t)^q \rangle$ exhibits multifractal scaling, 
$\langle P(r,t)^q \rangle \sim \langle P(r,t) \rangle^{\tau(q)}$, 
where $\tau(q) \sim q^\gamma$ and $\gamma=(d_w^\ell-1)/(d_w-1)$
and $d_w^{\ell}$ is the chemical random walk dimension.
This behavior is a consequence of the large fluctuations of $P(r,t)$ for fixed $r$ and $t$ from a given aggregate to another. 
Nevertheless, these authors have also shown that the distribution of the propagator in the chemical $\ell$ space, $P(\ell,t)$, is narrow and, consequently, $\langle P(\ell,t)^q \rangle \sim \langle P(\ell,t) \rangle^q$. The chemical distance $\ell$, the length of the shortest path between two sites along lattice bonds, is a more natural measure than the Euclidean distance in disordered systems. It is, for example, the distance used in the calculation of optimum paths in cities. 
 
Let ${\bf\ell}_{m,i}$ label the  $i$-th site out of those $n(m)$ that are placed at a chemical distance $\ell$ from a given origin with 
 $\ell_m\leq \ell < \ell_{m+1}$, $\ell_m=m\Delta \ell$ and $\Delta \ell$ small (say, of the order of the lattice spacing), and let $\Gamma_t({\bf\ell}_{m,i})$ be the survival probability in the chemical space defined as the probability  that site ${\bf\ell}_{m,i}$  has not been visited by time $t$ by a single random walker starting from the origin.
Then we can rewrite Eq.\ (\ref{SGSC1}) in the chemical $\ell$ space as \begin{equation}
\label{SGSCell}
S_N(t)=\sum_{m=0}^{\infty} \left\langle  \sum_{i=1}^{n(m)}\, 
\left\{ 1 - \left[\Gamma_t({\bf\ell}_{m,i})\right]^N \right\} \right\rangle\; .
\end{equation}
One may expect that the distribution of $\Gamma_t({\bf\ell}_{m,i})$ 
for fixed $\ell_m$ and $t$ is as narrow as the distribution of the propagator in the chemical space. (We support this conjecture in Sec.\ \ref{sect_3} by means of numerical simulations of $\Gamma_t({\bf\ell}_{m,i})$ in  two-dimensional percolation clusters at criticality.)
In this case
\begin{equation}
\Gamma_t({\bf\ell}_{m,i}) \simeq \langle \Gamma_t({\bf\ell}_{m,i})\rangle \equiv \Gamma_t(\ell_m)
\label{Gellaprox}
\end{equation}
for all possible lattice realizations so that 
$
\langle \left[\Gamma_t({\bf\ell}_{m,i})\right]^N\rangle \simeq
\langle \Gamma_t({\bf\ell}_{m,i})\rangle^N
$ 
and, therefore, $S_N(t)$ can be approximated by
\begin{equation}
\label{SGSC2ell}
S_N(t)= \sum_{m=0}^{\infty} \left\{1-\left[\Gamma_t(\ell_m)\right]^N \right\}\left\langle n(m) \right\rangle\; ,
\end{equation}
where $\left\langle n(m) \right\rangle$ is the average number of fractal sites separated from the origin by a chemical distance with value between  $\ell_m$ and $\ell_m+\Delta \ell$. 
From this formula and following the procedure outlined in the preceding paper \cite{fractal1}, in Sec.\ \ref{sect_2}  we will arrive at an expression for $S_N(t)$ for  the non-trivial time regime  whose leading asymptotic behavior coincides, apart from the value of the prefactor,  with the recent proposal, based on a scaling approach,  of Dr\"ager and Klafter \cite{DK}:
\begin{equation}
\label{DKsim}
S_N(t) \sim t^{d_s/2} (\ln N)^{d_\ell/v} 
\end{equation}
with $v=d_w^\ell/(d_w^\ell-1)$. Equation (\ref{DKsim}) differs from the relationship proposed by Havlin {\em et al.} \cite{Havlin}, Eq.\ (\ref{SNt1}), for those cases such as that considered in this present paper where $d_{\text{min}}\neq 1$.
Both Havlin {\em et al.} and Dr\"ager and Klafter supported 
their conjectures by means of data collapsing plots of computer simulation results obtained for two- and three-dimensional percolation aggregates, respectively. We here want also draw attention to the risk involved in this method of analysis when the influence of the corrective terms is not properly considered since these terms  have a large influence  \cite{usually} on the final value of $S_N(t)$. 

The paper is organized as follows.
In Sec.\ \ref{sect_2} we present the asymptotic evaluation of $S_N(t)$ for
stochastic fractal lattices as a translation to the chemical space of the procedure implemented in the Euclidean and deterministic fractal
cases \cite{fractal1,RC,PREeucl}. A less rigorous but fairly simple and instructive method for obtaining the main asymptotic term and estimating the corrective terms of $S_N(t)$ for large $N$ is also presented.
In Sec.\ \ref{sect_3} we report simulation results for the survival probability of a random walker on a two-dimensional incipient percolation aggregate when a trap is placed at a site at a fixed chemical distance $\ell$ or Euclidean distance $r$.
 We find that the distribution is  narrow [broad]  if the traps are located at a fixed  chemical [Euclidean] distance. 
The parameters governing (i) the asymptotic behavior of $ \Gamma_t(\ell)$ ($c$, $v$, $\mu$ and $A$), (ii) how the fractal volume grows ($V_0^{\ell}$ and $d_{\ell}$), and  (iii) how fast a single walker diffuses ($D_\ell$ and $d_w^{\ell}$) are estimated in this section.
In Sec.\ \ref{sect_4} we compare the zeroth- and first-order asymptotic expansion for $S_N(t)$ with simulation results obtained for the two-dimensional incipient percolation aggregate.  We also criticize the reliability of typical collapsing plots for the determination of the dominant trend of $S_N(t)$. We conclude with some remarks in Sec.\ \ref{sect_5}.

\section{TERRITORY COVERED BY $N$ RANDOM WALKERS ON A STOCHASTIC FRACTAL SUBSTRATE}
\label{sect_2}

In this section we will translate the results of the previous paper \cite{fractal1} to chemical language. 
Reasons for this procedure have already been given in
Sec.\ \ref{sect_1}. 
We start by replacing Eq.\ (\ref{SGSC2ell}) by its continuum 
approximation 
\begin{eqnarray}
\label{SNTCONT}
S_N(t)&\approx&\displaystyle\int_0^\infty\,\left\{ 1 - \left[\Gamma_t(\ell)\right]^N \right\}
 d_\ell\, V_0^{\ell}\, \ell^{d_\ell - 1} d \ell \; \nonumber \\
   &\approx& V_0^\ell d_\ell (2D_\ell)^{d_\ell/2} 
			t^{d_\ell/d_w} \int_0^\infty \left\{ 1 - \left[\Gamma_t(\xi)\right]^N \right\} \xi^{d_\ell - 1} d \xi ,
\end{eqnarray}
where $d V(\ell)=V_0^{\ell}\, d_\ell \, \ell^{d_\ell - 1} d \ell$ is the average number of fractal sites placed at a chemical distance between $\ell$ and $\ell+d\ell$, and $\xi\equiv\ell/(\sqrt{2D_\ell} t^{1/d_w^\ell})$. Here $D_\ell$ is the diffusion constant defined by the Einstein relation, 
\begin{equation}
\label{l2vst}
L^2= 2D_\ell t^{2/d_w^\ell},
\end{equation}
 where $L^2\equiv \langle \ell^2 \rangle$ is the mean-square chemical distance traveled by a single random walker by time $t$ ($t$ large).  
Next, we assume that the asymptotic dependence of $\Gamma_t(\ell)$ for  $\xi\gg 1$ is given by 
\begin{equation}
\Gamma_t(\ell) \approx
1-A \xi^{-\mu v} e^{- c \xi^{v}}
\left(1+\sum_{n=1}^\infty h_n \xi^{-n v}\right)\; ,
\label{gasin}
\end{equation}
with $v=d_w^{\ell}/(d_w^{\ell}-1)$. 
This functional form  holds (at least in its first terms) on Euclidean lattices \cite{RC,PREeucl,Larral1,Larral2} and agrees with the expression  conjectured in Refs.\ \cite{fractal1,RC} for fractal substrates with $d_{\text{min}}=1$ such as the Sierpinski gaskets. 
The dominant asymptotic behavior of the propagator in chemical space  \cite{SHDBA,RCBunde,Bunde},
 $P(l,t) \sim \exp\left(- c \xi^v \right)$, 
also coincides with the assumed dominant exponential decay of the mortality function $1-\Gamma_t(\ell)$ in Eq.\ (\ref{gasin}). 
It is known that both the propagator and the mortality function share the same asymptotic behavior for Euclidean lattices and for the Sierpinski lattice \cite{yusteacedoPRE} and we can expect that this coincidence also is the case for stochastic fractals (we will check this supposition in Sec.\ \ref{sect_3}). 
The rest of the analysis is identical with that carried out for Euclidean and deterministic fractal lattices  \cite{fractal1,RC,PREeucl} 
save for the change of the Euclidean parameters to their chemical analogues. The result for $S_N(t)$ is, consequently,
\begin{equation}
S_N(t)\approx \widehat{S}_N(t) \left(1-
\frac{d_\ell}{v} \displaystyle\sum_{n=1}^\infty\, \ln^{-n} 
N \, \displaystyle\sum_{m=0}^n \, s_m^{(n)} \ln^m \ln N \right)
\label{SNt}
\end{equation}
with 
\begin{equation}
\label{hatSNt}
\widehat{S}_N(t)=V_0^{\ell} (2D_\ell)^{d_\ell/2} t^{d_\ell/d_w^{\ell}} \left(\frac{\ln N}{c} \right)^{d_{\ell}/v} \; 
\end{equation}
and
\begin{eqnarray}
\label{s10}
s_0^{(1)}&=&-\omega  \\
s_1^{(1)}&=&\mu   \\
s_0^{(2)}&=&
-(\beta-1) \left( \frac{\pi^2}{12}+\frac{\omega^2}{2} \right) -
(c h_1-\mu \omega) \\
s_1^{(2)}&=& -\mu^2 + (\beta-1)\mu \omega \\
s_2^{(2)}&=& -\frac{1}{2} (\beta-1) \mu^2 \; .
\label{stilde}
\end{eqnarray}
Here $\omega=\gamma+\ln A+ \mu \ln c$, $\gamma \simeq 0.577215$ is the
Euler constant, and $\beta=d_\ell/ v$. 
The dependence on $t$ and $N$ of the main term of $S_N(t)$
as given by Eq.\ (\ref{hatSNt}), 
i.e., $S_N(t)\sim t^{d_s/2} (\ln N)^{d_{\ell}/v}$, coincides with the prediction of Ref.\ \cite{DK}.

\subsection*{A simpler way to estimate the territory covered}

\begin{figure}[t]
\begin{center}
\parbox{0.4\textwidth}{
\epsfxsize=\hsize \epsfbox{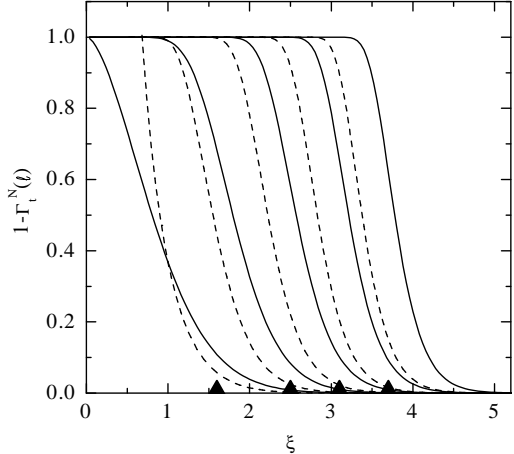}}
\caption{Function $1-[\Gamma_t(\ell)]^N$ versus $\xi=\ell/L(t)$ for (from left to right) $N=1, 10,100,1000$ and $ 10000$ where $\Gamma_t(\ell)=1-\xi^{-\mu v} \exp(-\xi^v)$, $c=1.0$, $v=1.7$, $\mu=0$ (solid line) and $\mu=0.8$ (dashed line). We have not plotted the unphysical values that appear in the case with $\mu=0.8$ when $\xi$ goes to zero. 
Notice the large influence of the subdominant power term $\xi^{-\mu v}$ on the value of $\Gamma_t(\ell)]^N$ which  will be reflected in the value of $S_N(t)$.
The triangles mark the value of $\xi_\times$ for (from left to right) $N=10,100,1000$ and  $10000$ obtained from Eq.\ (\protect\ref{xitimes0}). 
\label{figSTEP}}
\end{center}
\end{figure}

We finish this section by showing how to find the full main term of Eq.\ (\ref{SNt}) and even predict the form of the corrective terms by only resorting to extremely simple arguments.  The ideas of the following reasoning were already used in Ref.\ \cite{YL96}.
 The crucial point in our argument is that, for a  fixed time $t$, 
$1 - [\Gamma_t(\ell)]^N $ approaches a unit step function $\Theta(\ell-\ell_\times)$ when $N\rightarrow \infty$, $\ell_\times$ being the step's width (see Fig.\ \ref{figSTEP}). 
The reason for this behavior is clear: For large $N$, $[\Gamma_t(\ell)]^N$ is only non-neglible  when $\Gamma_t(\ell)$ is very close to 1.  Obviously this occurs when the root-mean square chemical distance $L(t)$ traveled by the single random walker by time $t$ is small compared with $\ell$, i.e., when $\xi=\ell/L(t)$ is large.  
This in turn implies that in the evaluation of $S_N(t)$ only the behavior of $\Gamma_t(\ell)$ for large $\xi$ is relevant. 
Then, as  $1 - [\Gamma_t(\ell)]^N $ approaches a step function of width $\ell_\times$, the integration of Eq.\ (\ref{SNTCONT}) yields
\begin{equation}
\label{SNtvolS}
S_N(t) \approx  V_0^\ell \ell_\times^{d_\ell},
\end{equation}
i.e., the territory covered is just the volume of a chemical hypersphere of radius $\ell_\times$. Defining the width $\ell_\times$ of the step function as the distance at which $1 - [\Gamma_t(\ell)]^N$ takes the intermediate value $1/2$ (any other value between 0 and 1 would also be valid as $\ell_\times$  is not very sensitive to this value when $N\gg 1$), and assuming that $1 - \Gamma_t(\ell)\approx A \xi^{-\mu v} \exp(-c \xi^v)$ for large $\xi$, we deduce that $ 1/2\approx N A \xi_\times^{-\mu v} \exp(-c \xi_\times^v)$, with $\xi_\times\equiv \ell_\times/L$ so that
\begin{equation}
c \xi_\times^v\approx \ln N-\mu v \ln \xi_\times+\ln2 A.
\label{elltimes1}
\end{equation}
The term $\ln N$ is  dominant on the right-hand side of Eq.\ (\ref{elltimes1}) for large $N$, so that a first-order solution of this equation is  
\begin{equation}
\label{xitimes0}
c \xi_\times^v\approx \ln N,
\end{equation}
i.e., $\ell_\times^v\approx L^v \ln(N)/c$. 
Hence  Eq.\ (\ref{SNtvolS}) yields 
\begin{equation}
S_N(t) \approx  V_0^\ell L^{d_\ell} \left(\frac{ \ln N}{c}\right)^{d_\ell/v}
\label{SNtsimple1}
\end{equation}
which is in full agreement with the main term of Eq.\ (\ref{SNt}) when the Einstein relation, Eq.\ (\ref{l2vst}), is considered. 
 Inserting the above first-order solution for $\ell_\times$ into the right-hand side of Eq.\ (\ref{elltimes1}), we get the improved solution 
$c \xi_\times^v\approx \ln N-\mu \ln\ln N +\ln A c^\mu+\ln 2$, so that Eq.\ (\ref{SNtvolS}) becomes 
\begin{eqnarray}
S_N(t) &\approx&  V_0^\ell L^{d_\ell} 
\left(\frac{ \ln N}{c}\right)^{d_\ell/v} \left[ 1+ \frac{d_\ell}{v}\; \frac{-\mu\ln \ln N + \ln A c^\mu+\ln 2}{\ln N} \right] .
\label{SNtsimple2}
\end{eqnarray}
This expression is strikingly close to the first-order approximation of  Eq.\ (\ref{SNt}), the only difference being that the term $\ln 2=0.693\cdots$ in Eq.\ (\ref{SNtsimple2}) plays the role of the Euler constant $\gamma\simeq 0.577215$ in Eq.\ (\ref{SNt}). 
Finally,  notice that  this simple method  is not limited to disordered media but that it is also valid for estimating  $S_N(t)$ for the non-disordered substrates (Euclidean and deterministic fractal media) considered in Refs.\ \cite{fractal1,RC,PREeucl}. 

\section{SURVIVAL PROBABILITY, FRACTAL VOLUME AND DIFFUSION IN PERCOLATION AGGREGATES}
\label{sect_3}
We have carried out simulations for the number of distinct sites visited
by $N$ independent random walkers on a typical stochastic fractal: the percolation aggregate embedded in two dimensions. 
In our simulations every random walker makes a jump from a site to one of its nearest neighbors placed at one unit distance in each unit time.
The incipient percolation aggregates were constructed by the standard Leath method \cite{Bunde,Leath} on a square lattice with side $400$. 
In the Leath method, a seed is placed on the site in the center of this box and the cluster is generated by epidemic spreading to their nearest neighbors with an infecting probability $p_c=0.5927460$ \cite{Bunde} corresponding to site percolation in the square lattice. 
At every step of the generation process a new chemical shell is added to the previous shell by the occupation with probability $p_c$ of its  empty nearest neighbors.
The process continues until we reach
a shell whose sites do not infect any of their empty neighbors; in this case the aggregate so generated is rejected.  If the
cluster grown in this way spans the box from a side to the other, the cluster is accepted as a good representation of a portion of an infinite percolation aggregate. Our simulations were carried out over 2000 aggregates generated in this way.

\begin{figure} 
\begin{center}
\parbox{0.4\textwidth}{
\epsfxsize=\hsize \epsfbox{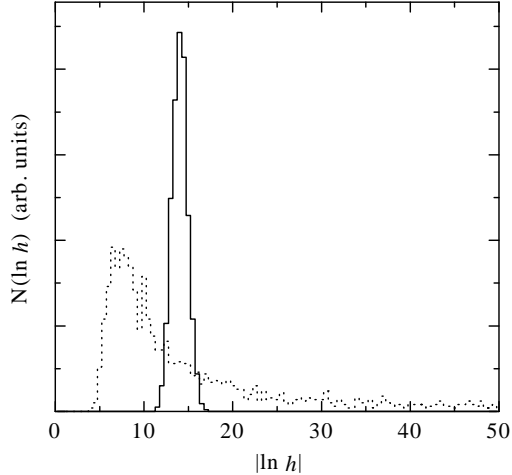}}
\caption{Plot of the histogram ${\cal N}(\ln h)$ versus $\vert \ln h \vert$ for fixed $r$ and $t$ (dotted line) and for fixed $\ell$ and $t$ (solid line). The values are $r=30$, $l=80$ and $t=1000$.\label{figHisto}}
\end{center}
\end{figure}

In order to compare the simulation results for $S_N(t)$ with the predictions of our theoretical approach,  Eq. (\ref{SNt}), we must check that the survival probability or, equivalently, the mortality function, $h(\ell,t)=1-\Gamma_t(\ell)$, really behaves  in the form conjectured in Eq. (\ref{gasin}). Moreover, we must confirm first that,  for a given chemical distance $\ell$, the distribution of $h(\ell,t)$ over different realizations of the incipient percolation cluster is narrow because our theoretical analysis [cf. Eq.\ (\ref{SGSC2ell})] relies on this assumption [cf. Eq.\ (\ref{Gellaprox})]. The numerical evaluation
of this quantity as well as the propagator $P({\bf i},t)$ (i.e., the probability of finding a single random walker at site ${\bf i}$ at time $t$)  is performed by the Chapman-Kolmogorov method  (also called the exact enumeration method \cite{SHDBA,Bunde}). Initially the propagator takes the value $1$ at the origin site, $P({\bf 0},0)=1$, and $0$ at any other site. This density evolves at every time step by
updating its value at every site in the form described by the master equation 
\begin{equation}
\label{CKupdate}
P({\bf i},t+1)=\frac{1}{b} \displaystyle\sum_{\text{neighbor}=1}^{M} \, P({\bf j},t)+\left(1-\frac{M}{b}\right) P({\bf i},t)\; ,
\end{equation}
where $b$ is the maximum coordination number ($b=4$ for the two-dimensional  percolation aggregate) and $M$ is the number of neighbors of site ${\bf i}$ that belong to the aggregate. 
This evolution equation is valid for the so-called blind ``ants'' \cite{SHDBA,Gennes} because the random walker ``attempts'' a jump at time $t+1$ to a possible nearest neighbor (selected randomly) of the site it occupied at time $t$. If the selected site does not belong to the cluster, the random walker stays at the same site and no jump takes place. The ``blindness'' of the random walkers is taken into account by the
second term of the right-hand side of Eq. (\ref{CKupdate}). 
The trap is simulated by a special site belonging to the cluster that absorbs all the probability density that enters it without giving back any probability to its neighbors. In the simulation of the mortality function, we located a trap at a chemical distance $\ell=30$ in each of the 2000 percolating clusters. 
We repeated the experiment for traps located at a fixed Euclidean distance, $r=80$. The resulting histogram for $t=1000$ is shown in Fig.\ \ref{figHisto} (to be compared with the histogram of the propagator shown in Fig. $4$ of Ref.\ \cite{RCBunde}).
One observes that the distribution corresponding to fixed $\ell$ is very narrow whereas the Euclidean version is broad and exhibits a long tail. 
The striking contrast between the two histograms in Fig.\ \ref{figHisto}
can be understood qualitatively. A fixed chemical distance $\ell$ between
the origin and the trap means that there exists at least a minimum path connecting those sites whose length is $\ell$. Thus, the minimum time taken by a random walker to arrive at the trap site is $t=\ell$ independently of the lattice structure. On the other hand, a fixed Euclidean distance $r$ could correspond to many different chemical distances from one cluster to another. This is a consequence of the stochastic lacunarity of the fractal aggregate. A large hole between the origin and the trap implies a large minimum time to travel around the border of that hole and, obviously, a small mortality in comparison with another cluster where no holes hinder the diffusion.

\begin{figure} 
\begin{center}
\parbox{0.4\textwidth}{
\epsfxsize=\hsize \epsfbox{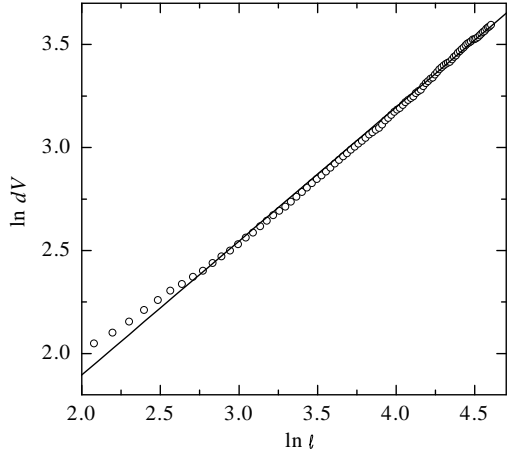}}
\caption{Plot of $\ln \langle \ell^2 \rangle$ versus $\ln t$,
 $\langle \ell^2 \rangle \equiv L^2$ being the chemical mean-square displacement of a single random walker calculated over $2000$ 
two-dimensional  percolation clusters at criticality.   
The line represents the function 
$L^2=2D_\ell \, t^{2/d_w^\ell}$ with $2D_\ell=1.2$ and $d_w^\ell=2.4$.\label{figEiRe}}
\end{center}
\end{figure}

Figure \ref{figEiRe} shows the chemical mean-square displacement 
\begin{equation}
L^2\equiv\langle \ell^2 \rangle=\left\langle \displaystyle\sum_{\text{sites}}\, \ell^2 P(\ell,t) \right\rangle \; 
\end{equation}
as a function of time. 
The propagator in the chemical space $P(\ell,t)$ is obtained by summing $P({\bf i},t)$ over all cluster sites ${\bf i}$ on the  chemical shell situated at distance $\ell$ from the origin. 
The result is compatible with the Einstein relation Eq.\ (\ref{l2vst}) with $2 D_{\ell}=1.20 \pm 0.1$ and $d_w^{\ell}=2.40 \pm 0.05$.   This value for $d_w^\ell$ coincides with that obtained in Ref.\ \cite{RCBunde} and is in agreement with the value reported in Refs.\ \cite{SHDBA,Majid}. 

\begin{figure} 
\begin{center}
\parbox{0.4\textwidth}{
\epsfxsize=\hsize \epsfbox{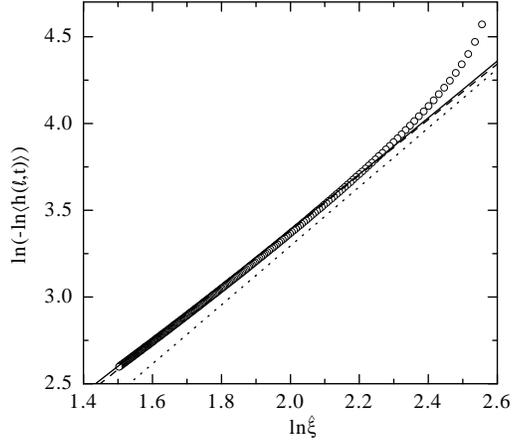}}
\caption{ Plot of $\ln(-\ln h(\ell,t))$ versus $\ln \hat \xi$ averaged over $2000$ two-dimensional incipient percolation clusters. The trap was always placed at a site at distance $\ell=80$ from the origin. 
The lines represent the functions 
$h(\ell,t) = \hat A \hat \xi^{-\mu v} \exp(-\hat c \hat \xi^v) $
where $\{ \hat A = 0.9$, $\mu = 0.8$, $\hat{c} = 0.9$, $v = 1.7\}$ (solid line), 
$\{\hat A = 1$, $\mu = 0$, $\hat c = 1.2$, $v = 1.6 \}$ (dashed line), and 
$\{ \hat A = 1$, $\mu = 0$, $\hat c = 0.9$, $v = 1.7 \}$  (dotted line). 
\label{figlnlnh}}
\end{center}
\end{figure}

 In Fig.\ \ref{figlnlnh} we plot
$\ln(-\ln  h(\ell,t))$
 versus 
$\hat \xi\equiv \ell/t^{1/d_w^{\ell}}$ 
with $\ell=80$ and, according to the previous discussion, $d_w^{\ell}=2.40$. 
If the conjecture in Eq.\ (\ref{gasin}) is right, we can take $h(\ell,t)\sim \exp(-\hat c \hat \xi^v)$ as a first approximation, and hence should observe the linear behavior $\ln(-\ln h(\ell,t)) \sim \ln \hat c+v \hat \xi$ with 
$\hat \xi =\sqrt{2D_\ell} \xi$ and $\hat c=c/(2D_\ell)^{v/2}$.
Certainly the plot seems linear except for a portion in the range $\hat \xi \gtrsim 2.2$. 
This is a finite size effect (already analyzed in the case of the two-dimensional Sierpinski gasket in Ref.\  \cite{yusteacedoPRE}) associated with the existence of a minimum arrival time corresponding to a random walker who travels ``ballistically'' along a chemical path from the origin to the trap, which in turn implies a maximum available value of $\hat \xi$ in the simulations (in our simulations this maximum value is $80/80^{1/d_w^\ell}\simeq 12.9$). A reliable interval for numerical fits should exclude this very short time regime. A linear fit in the interval $1.6 \le \ln \hat \xi \le 2.17$, corresponding to $200\le t\le 1000$, gives the values $\hat c=1.2\pm0.1$, i.e., $c=1.3\pm0.1$, and $v=1.6 \pm 0.05$. The dashed line in Fig.\ \ref{figlnlnh} corresponds to these values.
The good agreement with numerical values in the above interval seems to assure the correctness of the approximation $h(\ell,t)\sim h_{\text{a}}(\ell,t)=\exp(-\hat c \hat \xi^v)$ with the values of $c$ and $v$ given above. 
However, the solid line in Fig.\ \ref{figlnlnh} is a challenge to this interpretation: one sees that the function $h_{\text{b}}(\ell,t)= \hat A \hat \xi^{-\mu v} \exp(-\hat c \hat \xi^v)$ with $v=1.70$, $\hat c=0.9$ (i.e. $c=1.05$), $\mu=0.8$ and $\hat A=1.1$ is as good as $h_{\text{a}}(\ell,t)$. Indeed, $h_{\text{b}}(\ell,t)$ is more consistent from a theoretical point of view  than $h_{\text{a}}(\ell,t)$  because the expected theoretical value of $v$  corresponding to $d_w^\ell=2.40$  is $v=d_w^\ell/(d_w^\ell-1)=1.71$,  which is in better agreement with the exponent $v=1.7$ of $h_{\text{a}}(\ell,t)$ than with the exponent $v=1.6$ of  $h_{\text{b}}(\ell,t)$. 
 Finally, it should be noticed that the values $c=1.05$, $v=1.70$  are also in agreement with the corresponding parameter values of the propagator \cite{RCBunde}, thus supporting the guess made in Sec.\ \ref{sect_2} [see below Eq.\ (\ref{gasin})] that the dominant exponential term of the propagator and of the mortality function are the same. 
This leads us to consider that the set of parameters ${\hat c=0.9, v=1.7, \mu=0.8}$ is more reliable than  ${\hat c=1.2, v=1.6, \mu=0}$.
Obviously, further intensive (and extremely time consuming) computer simulations for the mortality function would be required in order to reliably determine  the values of the parameters that appear in Eq.\ (\ref{gasin}) and in the asymptotic corrections of $S_N(t)$.

\begin{figure} 
\begin{center}
\parbox{0.4\textwidth}{
\epsfxsize=\hsize \epsfbox{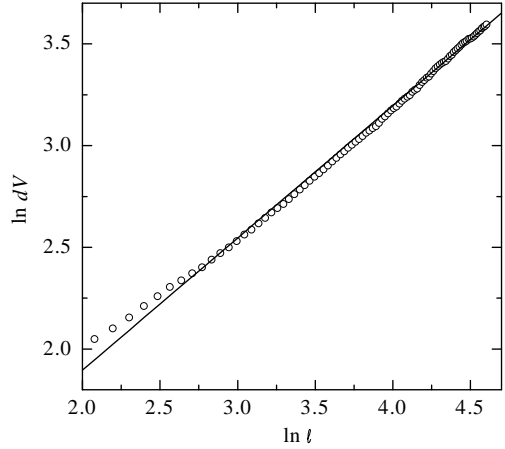}}
\caption{ Plot of  $\ln dV(\ell)$ versus $\ln \ell$ where  
$dV(\ell) = V(\ell + 1) - V(\ell)$ is the average number of sites in the chemical shell  at distance $\ell$. 
The line represents the function $V_0^\ell d_\ell \ell^{d_\ell-1}$ with $V_0^\ell=1.1$ and $d_\ell=1.65$. \label{figVolQ}}
\end{center}
\end{figure}

We have also evaluated numerically the fractal volume in terms of the
chemical distance $V(\ell)$, i.e., the number of lattice sites inside a
circumference (in chemical space) of radius $\ell$.  The results are shown in Fig.\ \ref{figVolQ}.
A good fit to 
$dV(\ell)=d_\ell V_0^{\ell} \ell^{d_\ell-1} d\ell$ 
is found with $V_0^{\ell}=1.1\pm0.2$ and $d_\ell=1.65\pm0.05$. 
Taking into account that $d_f=91/48$, we deduce that $d_{\text{min}}=d_f/d_\ell=1.15 \pm 0.05$,  which agrees with previous estimates  \cite{SHDBA,Bunde}. 

\section{SIMULATION RESULTS: TERRITORY COVERED BY $N$ RANDOM WALKERS ON
THE PERCOLATION AGGREGATE}
\label{sect_4}
As discussed in Sec.\ \ref{sect_1}, the average implicit in the evaluation of $S_N(t)$ is double: first, we take an average over experiments performed on the same aggregate and then a second average over different percolation clusters.
The $N$ random walkers are always placed initially upon the site in the center of the square box. 
In our simulations we have performed an average over $100$ runs per cluster and a second average over $2000$ percolation clusters in order to achieve  good statistics. The maximum time considered was $t=1000$. 

\begin{figure} 
\begin{center}
\parbox{0.4\textwidth}{
\epsfxsize=\hsize \epsfbox{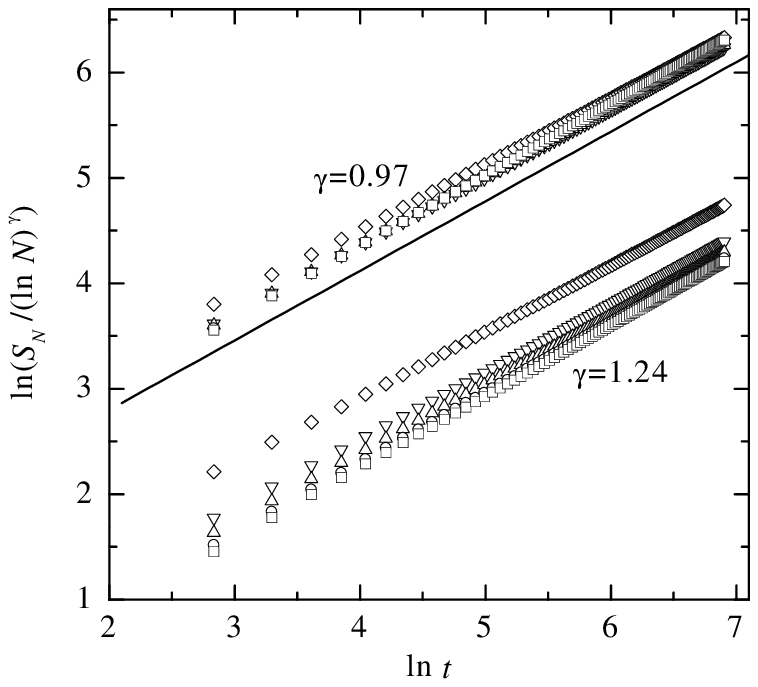}}
\caption{Plot of $\ln(S_N(t)/(\ln N)^{\gamma})$ versus $\ln t$ for $N=2^2$ (diamond), $2^5$ (down triangle), $2^8$ (up triangle), $2^{11}$ (circle), and $2^{13}$ (square) in the two-dimesional  percolation aggregate with  $\gamma=d_\ell/v\simeq 0.97$  and $\gamma=d_f/u \simeq 1.24$. 
The values corresponding to $\gamma=d_\ell/v\simeq 0.97$ have been shifted up by 3/2.  The line has a slope equal to $d_s/2\simeq 0.66$
\label{figColapsoPerco}}
\end{center}
\end{figure}

According to Eqs.\ (\ref{SNt}) and  (\ref{hatSNt}), the quotient
$S_N(t)/(\ln N)^{\gamma}$
with $\gamma=d_\ell/v$ is only a function of $t$.
In Fig.\ \ref{figColapsoPerco} the logarithm of that quotient
is plotted versus $\ln t$ for several values of $N$.
The data collapse  and the slope close to $0.66$ seems to support Eq.\ (\ref{DKsim}) with $\gamma=d_\ell/v=0.97$, which is in agreement with similar recent results for the three-dimensional percolation aggregate \cite{DK}. 
The collapse is, however, slightly poorer when the exponent $\gamma=d_f/u=1.24$  ($d_f=91/48$ and $d_w=2.87$ \cite{SHDBA,Bunde}) proposed by Havlin {\em et al.} \cite{Havlin} [see Eq.\ (\ref{SNt1})] is  used, as Fig.\ \ref{figColapsoPerco} shows. 
So, one migth be led to the conclusion that the
correct value of $\gamma$ as defined above is $d_\ell/v$.
But in this analysis there was no consideration of the relatively large logarithmic corrections predicted by the asymptotic analysis presented in Section \ref{sect_2}, so that the reliability of the above conclusion is seriously affected by this omission. 

\begin{figure} 
\begin{center}
\parbox{0.4\textwidth}{
\epsfxsize=\hsize \epsfbox{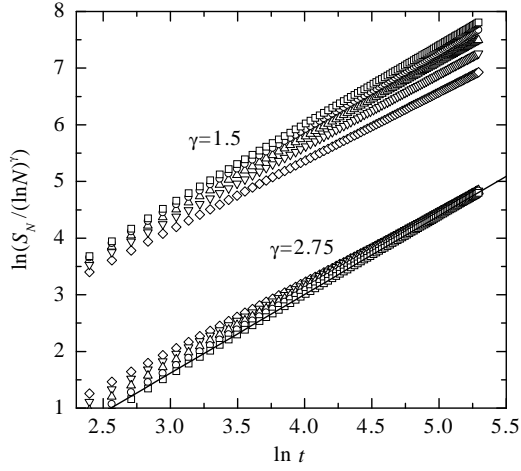}}
\caption{Plot of $\ln(S_N(t)/(\ln N)^{\gamma})$ versus $\ln t$ with 
$N=2^8$ (diamond), $2^{10}$ (down triangle), $2^{12}$ (up triangle), $2^{14}$ (circle), $2^{16}$ (square) for the three-dimensional Euclidean lattice with $\gamma=3/2$  and $\gamma=2.75$. 
The line has a slope equal to $1.4$.
\label{figColapsoEu3d}}
\end{center}
\end{figure}

To illustrate this point, let us now carry out the same kind of analysis with the simulation results of $S_N(t)$ when the substrate is a three-dimensional Euclidean lattice. For this case it is well known \cite{RC,PREeucl} that $S_N(t)$ is given by an asymptotic expression with the form  of Eq.\ (\ref{SNt}) in which the logarithmic corrective terms are very important even for very large values of $N$. 
Indeed, the main asymptotic term leads to very poor predictions for 
$S_N(t)$, whereas  the second-order approximation ($n=2$) gives excellent agreement with numerical simulation results. 
The exponent $\gamma$ of the main logarithmic term in $N$  and the time exponent $d_\ell/d_w^\ell=d_s/2$ are equal to $3/2$.
We have plotted in Fig.\ \ref{figColapsoEu3d} the quotient $S_N(t)/\ln^\gamma N$ versus $\ln t$ for several values of $N$  taking into account that the rigorous value of $\gamma$ is $3/2$. We see that the collapse is far from being perfect because the logarithmic corrections have been ignored. 
Nevertheless, an effective (but incorrect!) value of $\gamma=2.75$  yields
a much better data collapse and a slope close to the theoretical value $d_s/2=1.5$. 
We thus conclude that  analysis of data collapse plots based on the form of the main term of quantities such as $S_N(t)$ (which typically exhibit large corrective terms) should be performed with caution.  The values of the exponents estimated in this way are untrustworthy because the existence of logarithmic corrections to the main term cannot simply be ignored. 
The value of $\gamma=2.75$  obtained before is then only an effective way of including all these corrective terms together but the true expression involves a main term of the form $(t \ln N)^{3/2}$ times a series similar to that given in Eq.\ (\ref{SNt}). These considerations should prevent us from drawing hasty conclusions from a simple view of plots such as Figs.\ \ref{figColapsoPerco} and  \ref{figColapsoEu3d}.

\begin{figure} 
\begin{center}
\parbox{0.4\textwidth}{
\epsfxsize=\hsize \epsfbox{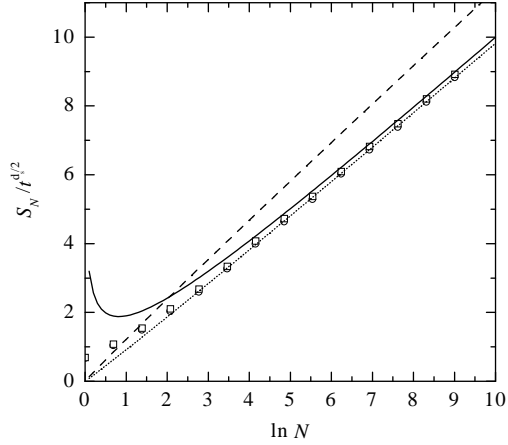}}
\caption{Plot of  $S_N(t)/t^{d_{\ell}/d_w^{\ell}}$ versus $\ln N$  in
the two-dimensional incipient percolation aggregate for $N=2^0,2^1,\ldots,2^{13}$. The circles [squares] are the simulation results for $t=1000$ [$t=500$] averaging over $2000$ aggregate realizations. 
The dashed [dotted]  line is the zeroth-order theoretical prediction with $c=1.05$ and $v=1.7$ [$c=1.3$ and $v=1.60$] and the solid line is the first-order approximation with $c=1.05$, $v=1.7$, $\mu=0.8$ and $A=1$. 
\label{figSNt}}
\end{center}
\end{figure}

Finally, in Fig.\ \ref{figSNt} we show the dependence of $S_N(t)$ on $N$ and compare simulation results with the zeroth- and first-order asymptotic prediction given by Eq.\ (\ref{SNt}). 
When the parameter set ${\hat c=0.9, v=1.7, \mu=0.8, A=1}$ (see Sec.\  \ref{sect_3}) is used,  we get results with a very familiar aspect as they are quite similar (although, perhaps the first-order approximation is too good) to that already found for Euclidean \cite{RC,PREeucl} and Sierpinski lattices \cite{fractal1}. This is indeed encouraging. 
However, when the parameters $\hat c=1.3$ and $ v=1.6$ are used, we obtain a surprising and strikingly accurate zeroth-order approximation.
At this point, we again suspect that this last set of parameters are only effective parameters that include the influence of the true logarithmic corrective terms in the range of $N$ simulated. 
Hence, Fig.\ \ref{figSNt} illustrates again, but from a different perspective, how the omission of important corrective terms could lead to finding effective parameters that, although providing excellent approximations in the (relatively short) range under consideration, are really erroneous. 

\section{Summary}
\label{sect_5}

In this paper, the average
fractal territory covered up to time $t$ by $N$ independent random walkers   all starting from the same origin on stochastic fractal lattices is calculated in terms of an asymptotic series expansion,
$\sum_{n=0}^\infty\sum_{m=0}^n s_{nm}(\ln N)^{d_\ell/v-n} (\ln\ln N)^m$ [see Eq.\ (\ref{SNt})], 
 which is formally identical to those obtained for Euclidean and deterministic fractal lattices. 
Equation (\ref{SNt}) is obtained by assuming that (i)  the average fractal volume inside a ``hypersphere'' of chemical radius $r$ grows as $V_0^{\ell} r^{d_\ell}$, (ii) the  distribution of the the short-time survival probability of a single random walker in the presence of a trap is narrow, so that Eq.\ (\ref{Gellaprox}) holds, and (iii) this short-time survival probability is asymptotically given by Eq.\ (\ref{gasin}). 
We performed numerical simulations for the two-dimensional percolation aggregate at criticality which support the validity of the above assumptions.
The  zeroth- and first-order theoretical asymptotic expression of $S_N(t)$ were calculated explicitly and, for the two-dimensional percolation aggregate, they compared reasonably well with numerical simulation results.
The agreement is similar to that found for non-disordered media.

In our procedure, the use of the chemical distance turns out to be fundamental because the distribution of the short-time survival probability    in the chemical space is so narrow that we can safely replace  the power $N$ of the mean value of the survival probability by the mean value of the power $N$ of the survival probability. (On the contrary, this does not hold at all when Euclidean distances are used.) 
This allowed us to easily translate the theoretical results previously derived for Euclidean and deterministic fractals \cite{fractal1,RC,PREeucl} to
disordered media.

\acknowledgments
This work has been supported by the DGICYT (Spain) through Grant No. PB97-1501 and by the Junta de Extremadura-Fondo Social Europeo through Grant No. IPR99C031. 



\begin{references}
\bibitem{usually} At least, this is true for one-, two- and three-dimensional Euclidean lattices \protect\cite{RC,PREeucl} and for the two-dimensional Sierpinski gasket \protect\cite{fractal1}.
\end{references}
\end{document}